\documentclass[fleqn]{article}
\usepackage{espcrc2}
\usepackage{epsfig}

\newcommand{\fslash}[1]{\slash\hspace{-.068in}#1}
\newcommand{\Dslash}{\slash\hspace{-.098in}D}
\newcommand{\projectree}{{\cal P}(v)}
\title{Four-momentum boosted Fermion fields}

\author{
P.A.~Boyle\address[ED]{Department of Physics and Astronomy, University 
  of Edinburgh, Edinburgh EH9 3JZ, Scotland}\address[Columbia]{Department 
  of Physics, Columbia University, New York, NY, 10027}
}     

\begin{document}
\begin{abstract}
A formulation of the fermion action is discussed which includes
an explicit four momentum boost on the field prior to discretisation.
This is used to shift the zero of lattice momentum to lie near one
of the on-shell quark poles. The positive pole is selected if we wish to 
describe a valence quark, and negative pole for a valence anti-quark. 
Like NRQCD, the typical lattice momenta involved in hadronic correlation 
functions can be kept small:
of order $O(a \Lambda_{\rm QCD})$, rather than $O(a m_Q)$ even when describing 
heavy quarks. 
If we expand around the particle pole, the
anti-particle correlator will be poorly described for large $a m_Q$.
However, in that case the anti-particle will be far off shell and will
only affect unphysical, renormalization factors.
The formulation produces the correct continuum limit, and preliminary 
results have been obtained (for an unimproved action) of both the 
one-loop self energy and a non-perturbative correlation function.
\end{abstract}
\maketitle
\vspace{-0.1in}
\section*{Introduction}
\vspace{-.1in}
The heavy quark effective theory can be derived from the Minkowski 
path integral in two stages \cite{Mannel}. 
Firstly a change of variables 
$
\psi^\prime = e^{i q\cdot x} \psi
$
is made on the path integral to include an explict four-momentum
$q_\mu = m v_\mu$ in the quark fields, to boost 
to the rest-frame $v_\mu$ of the hadron:
\begin{equation}
G^n(x_1,\ldots,x_n) =
\int d\bar\psi d\psi \bar\psi(x_1)\ldots\psi(x_n) e^{i S_F}
\label{labEq1}
\end{equation}
\begin{equation}
G^n = e^{i q\cdot x_1}\ldots e^{-i q\cdot x_n}
\tilde{G}^n(x_1,\ldots,x_n)
\end{equation}
\begin{equation}
\tilde{G}^n = 
\int d\bar\psi^\prime d\psi^\prime
\bar\psi^\prime(x_1)\ldots\psi^\prime(x_n) e^{i S_F^\prime}
\label{labEq2}
\end{equation}
where we can immediately perform the derivative 
that pulls down the boosting phase:
\begin{equation}
S_F^\prime =
\int d^4z \bar\psi^\prime(z)
\left[ i \Dslash + m + \fslash{q} \right]\psi^\prime(z)
\label{EqContAction}
\end{equation}
To tree level we can immediately identify an anti-particle projected 
mass term as $2 m \projectree$ where the projector 
$\projectree = (1+\fslash{v})/2$. Beyond tree level the boost
to the frame $v_\mu$ will contain counterterms that keep
the renormalised boost in the desired location
and we write the bare mass term as $m + \fslash{q}(g,m,v)$.

We may choose to formulate a discretised theory from either representation 
of the path integral \ref{labEq1} or \ref{labEq2}.
However, we will argue that the latter
is the preferable starting point for a lattice action for hadrons 
containing heavy quarks. 

In  HQET and (moving) NRQCD \cite{NRQCD,MNRQCD} a Foldy-Wouthuysen 
transformation is performed on the four-spinors, 
and the lower component integrated out.
In NRQCD a truncated expansion in $\frac{1}{a M_Q}$ is performed on the 
resulting  effective action, and loop corrections to this expansion 
become divergent as the quark mass becomes less than the inverse lattice 
spacing. As noted in \cite{FermilabMFLGT}, 
by retaining lower components, we maintain the attractive 
property of possessing a continuum limit.
\vspace{-0.1in}
\section*{Formulation}
\vspace{-.1in}
The action (\ref{EqContAction}) can be Euclideanised and then
discretised using any of the currently used four-component lattice 
Dirac operators, which we label generically as $D^L$. 
This generates a Euclidean lattice action for the boosted field,
where we take the Euclidean momentum $q^E_4 = i q_0^M$
$$
S^E_F = \sum\limits_z \bar\psi(z) \left[ D^L +   m_0 + i \fslash{q}^E(g,m_0,v) \right] \psi(z)
$$
This action has a modified mass term containing a particle 
(anti-particle if $q_0^M < 0$)
projector, and is exactly analogous to HQET where the particle fields
are slowly varying, and the anti-particle fields acquire a doubled mass
term.
\vspace{-0.1in}
\section*{Tree level masses}
\vspace{-.1in}
It is simple to write down the Feynman rules for our action in terms
of the Feynman rules for the massless action $\bar\psi D^L \psi$.

We denote the Feynman rules associated with the lattice derivative $D^L$
as propagator $G^L(p)^{-1} = i \gamma_\mu S_\mu(p) + R(p)$, and 
quark n-gluon vertices $V^n(p_q,p_q^\prime,k_{g1},\ldots,k_{gn})$.
The quark-gluon vertices are unaltered for the boosted action, 
the modified propagator is given by:
$$
\tilde{G}^\prime(p)^{-1} = i \gamma_\mu S_\mu(p) + R(p) + m_0 + i \fslash{q}^E(g,m_0,v)
$$
For the Wilson action, $S_\mu(p) = \sin p_\mu$ and R(p) encodes the
Wilson term. Generically, the correct continuum limit requires
$$
\begin{array}{c}
S_\mu(p) = p_\mu + O(p^3)\\
R_(p) = R_{\mu\nu} p_\mu p_\nu + h. o.\\
\end{array}
$$
While for an $O(a)$ off-shell improved derivative we require of the
chiral symmetry breaking term $R_{\mu\nu}= 0$,
the clover action has $R_{\mu\nu} = \frac{1}{2} \delta_{\mu\nu}$ and 
applies the equation of motion to ``redefine'' the
mass at order $(am)^2$. This trick is not appropriate here due to the 
Dirac structure and size of our mass term.

The low momentum expansion of the denominator of the $\tilde{G}$ propagator for
the tree level boost $q^E_4 = i m_0$, and $q_j = 0$ is:
\begin{equation}
- 2 i m_0 p_0 + \vec{p}^2 + 2 m_0 R_{ij} p_i p_j
\end{equation}
This low momentum expansion immediately
implies the solutions for the pole ($m_1$) and kinetic ($m_2$) masses in 
Table~\ref{TabDispersion}.

\begin{table}[thb]
\vspace{-.2in}
\caption{\label{TabDispersion}
Masses in the low order dispersion relation
}
\begin{tabular}{c|c|c|c}
  &$ m_1$ &$ m_2(R=0)$ & $ m_2( D^W )$\\
\hline
$\tilde{G}$ & $0$   & $m_0 $& $\frac{m_0}{1+m_0}$\\
${G}$       &$ m_0$ & $m_0$ & $\frac{m_0}{1+m_0}$\\
{\tiny Wilson} &$\log (1 + m_0)$ & N/A &$e^{m_1} \frac{\sinh m_1}{1+\sinh m_1}$ 
\end{tabular}
\vspace{-.2in}
\end{table}
Thus the boosted Fermion approach automatically produces
the perfect $O(p^2)$ dispersion relation at tree-level, unlike other four
component approaches, \emph{provided} the lattice derivative used is O(a) improved.
This is not the case for the Wilson derivative, but would certainly be
the case for a Neuberger or Domain Wall based lattice operator.
Alternative schemes have also been proposed for removing the doublers which
introduce no such classical O(a) term \cite{PaulMackenzie}. Regardless,
in what follows we treat the boosted Wilson case as an expedient
toy model for testing out the approach.
\vspace{-.1in}
\section*{Momentum dependence of vertices}
\vspace{-.1in}
To higher order, in calculations of vertex function renormalisation or the
self energy, for example, we are interested in the \emph{on-shell} 
point for external quark momenta. 
For standard four component formulations
this introduces a large imaginary temporal 
lattice momenta $q_4 \simeq i m_1$, which 
flows through and distorts the lattice quark gluon vertices
as shown in Table~\ref{TabVertices}. 
\begin{table*}[thb]
\caption{\label{TabVertices}
Mass dependence of temporal gluon coupling 
in typical lattice Feynman graphs
}
\begin{tabular}{c|c|c}
& Boosted Wilson & Heavy Wilson \\
\hline
$qqg_t$
& $-ig \left[ \gamma_0 \cos \frac{1}{2}k_0 - i r \sin \frac{1}{2}k_0 \right]$
& $-ig \left[ \gamma_0 \cosh (m_1 - i \frac{1}{2}k_0) - i r \sinh (m_1-i\frac{1}{2}k_\mu) \right]$ \\
$qqgg_t$
&$ -g^2 \left[ r \cos \frac{1}{2}k_0 - i\gamma_0 \sin \frac{1}{2}k_0 \right] $
&$ -g^2 \left[ r \cosh (m_1-i\frac{1}{2}k_0) - i\gamma_0 \sinh (m_1 -i\frac{1}{2}k_0) \right] $
\end{tabular}
\vspace{-.15in}
\end{table*}
Indeed this mechanism is responsible for
the dominance of the temporal gluon vertex that renders the perturbative
equivalence of the heavy Wilson action with the static Wilson line.
In contrast, the on-shell point of the boosted approach
is by construction at the zero of external lattice momentum,
and the quark gluon couplings are minimally distorted.
\vspace{-.1in}
\section*{One loop mass renormalisation for $D^{\rm Wilson}$}
\vspace{-0.1in}
We have calculated preliminary results for the one-loop self
energy with both boosted and regular Wilson fermions.
The result for regular Wilson fermions agrees with those 
of \cite{KronfeldMertens}. In the Wilson case
the one-loop self energy $\Sigma^{[1]}$
(as opposed to mass correction $m_1^{[1]}$) grows exponentially
with the pole mass via the couplings in Table~\ref{TabVertices}.
For boosted Wilson Fermions, however,
the zero external  four-momentum leaves the
tadpole graph independent of the mass, and the
the rainbow graph is only weakly dependent on the 
quark mass. A table of these results will be presented in a subsequent
publication~\cite{InPrep};
\vspace{-.1in}
\section*{Non-perturbative correlator with $D^{\rm Wilson}$}
\vspace{-0.1in}

For our toy $D^W$ case, the correct massless limit occurs
at $m_0 = m_{\rm crit} \ne 0$. 
The required boost $q_4$ should vanish when and $\vec{q}_j = 0$ and 
$m_0 = m_{\rm crit}$.
We therefore input the known non-perturbative $m_{\rm crit}$ and take
$
m_0 = \bar{m} +  m_{\rm crit}$, 
$
q_4 = i(\bar{m} - \Lambda$), 
$q_j = 0$.

This subtraction of $m_{\rm crit}$ is a technicality that
is unique to the Wilson Dirac operator.

If the behaviour of $q(m,g,v)$ is benign
we will not have a significant ``tuning'' problem to locate the 
required non-perturbative boost, and  
then a massless mode will arise for small values 
of $\Lambda$. In fact, it proved necessary to not make the regulating
parameter $\Lambda$ too small in order to control the expense 
of solving the propagator. 

\begin{figure}[hbt]
\vspace{-.5in}
\caption{\label{FigCorr} 
Heavy-light $Qq$ pseudo-scalar correlator. 
$\bar{m} = 2.0$, $\Lambda = 0.1$ and $\kappa_q = 0.1550$
at $\beta = 6.0$ with Wilson Dirac operator for both light and heavy quarks.
The forward propagator is light, and the 
heavy anti-particle mode is clearly seen propagating backwards.
}
\vspace{-.1in}
\epsfig{file=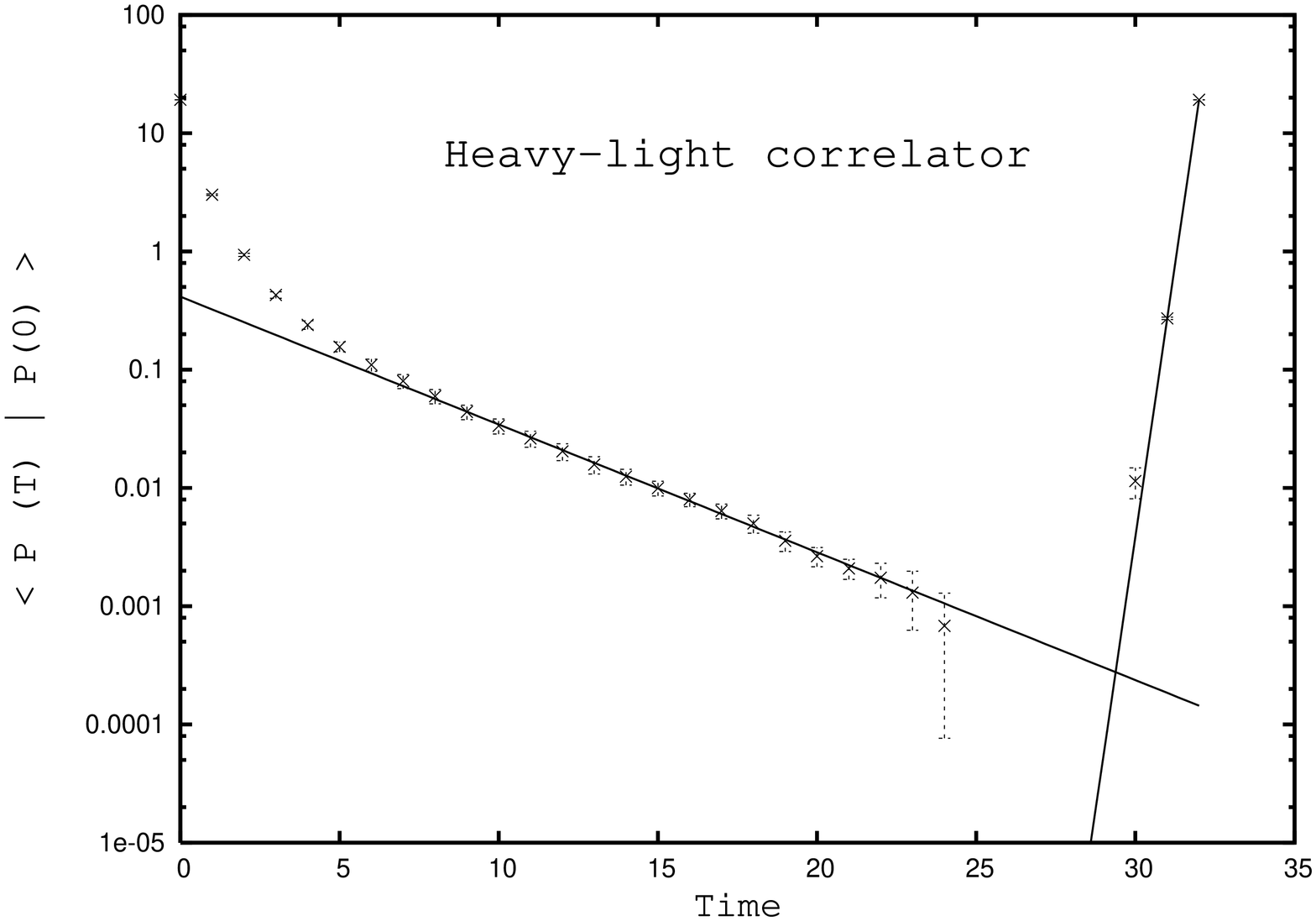,width=5cm,angle=270}
\vspace{-.25in}
\end{figure}

The solver for the projected mass term 
is simple to implement, and preconditions easily.
Some inversions have been performed on a $16^3\times 32$ 
quenched lattice at $\beta = 6.0$, and a heavy-light pseudoscalar 
correlation function obtained in Figure~\ref{FigCorr}. 
\vspace{-.3in}
\section*{Conclusions}
\vspace{-.1in}
While the data are not relevant for physical conclusions, the features of
the correlation function successfully demonstrate the method.
The light forward propagating particle mode is clearly present,
with $aM_P^+ \simeq 0.25$, while the heavy backwards propagating
anti-particle mode is also clearly seen, with $aM_P^- \simeq 4.0$.

When the technique is coupled with an $O(a)$ off-shell improved
Dirac operator such as $D^{\rm Neuberger}$ it will
lead to much improved lattice treatment of the
semi-relativistic regime, such as charmonium and D mesons.

The ability to insert spatial momenta in the action is highly 
promising for calculations with difficult kinematics, such
semi-leptonic decays.
\vspace{-.1in}
\section*{Acknowledgements}
\vspace{-.1in}
During the lattice conference I became aware that a similar
four-momentum boosted approach has been independently pursued by Paul
Mackenzie \cite{PaulMackenzie,InPrep}, and I thank him for a copy of his 
unpublished notes. 
The Mackenzie approach differs significantly in the technique for improvement 
and avoiding Fermion doubling.

I thank Norman Christ for useful conversations and Norman Christ,
Paul Mackenzie and Huey-Wen Lin for critical readings of the manuscript.

Configurations were kindly provided by the RBC collaboration,
and all code was written within the Columbia Physics System.

\end{document}